# Joint measurement of cosmic-ray muons and seismic waves at laboratory scale

Jun Matsushima,[1] Masashi Kodama,[1,2] Mohammed Y. Ali,[3] Fateh Bouchaala,[3] Hiroyuki K.M. Tanaka,[4] Tadahiro Kin,[5] Hamid Basiri,[1] Toshiyuki Yokota[2] and Makoto Suzuki[6]

[1]*Graduate School of Frontier Sciences, The University of Tokyo, 277-8563, Chiba, Japan, E-mail: jun-matsushima@edu.k.u-tokyo.ac.jp*
[2]*Research Institute for Geo-Resources and Environment, National Institute of Advanced Industrial Science and Technology (AIST), 305-8560, Ibaraki, Japan , E-mail: kodama-ms996@aist.go.jp*
[3]*Department of Earth Sciences, Khalifa University of Science and Technology, P.O. Box: 127788, Abu Dhabi, UAE*
[4]*Earthquake Research Institute, The University of Tokyo, 113-0032, Tokyo, Japan*
[5]*Interdisciplinary Graduate School of Engineering Sciences, Kyushu University, 816-8580, Fukuoka, Japan*
[6]*Graduate School of Engineering, The University of Tokyo, 113-8656, Tokyo, Japan*

## SUMMARY

Current geophysical exploration methods face challenges in accurately determining gas saturation levels and elastic constants with adequate spatial resolution. Seismic wave velocity is a critical physical property in these techniques, but it introduces uncertainties because of its composite nature involving density and two elastic constants (e.g. bulk and shear modulus), which exhibit a trade-off relationship. We propose a novel approach that integrates cosmic-ray muon detection with seismic exploration to independently resolve *P*- and *S*-wave velocities into their constituent elastic constants and densities. First, we utilized a fluid substitution approach based on Gassmann's model to illustrate the benefits of incorporating density information in predicting gas saturation levels in pores. This supports the advantage of decomposing seismic wave velocity into density and two elastic constants. Second, to validate the applicability and performance of the proposed method, which involves separating seismic wave velocity into density and two types of elastic constants, muon and ultrasonic data were collected in laboratory experiments on two different targets: an acrylic block and an aluminium block. Upon muon observation, a relationship is established to convert muon flux into density length, considering the characteristics of the building housing the laboratory and the direction of muon arrival at specific positions within the building. Although there is potential for enhancing the accuracy of the derived physical properties such as density, bulk modulus, and shear modulus, the feasibility of this method has been successfully demonstrated at the laboratory scale.

## 1. INTRODUCTION

The seismic reflection method is an essential tool for investigating subsurface geology, supporting a broad spectrum of applications in natural resource exploration, engineering, and environmental management. Time-lapse seismic techniques are extensively utilized in various geophysical surveys, including monitoring oil and gas fields (Lumley 2001; Campbell *et al.* 2011; Oghenekohwo *et al.* 2017), $CO_2$ sequestration (Daley *et al.* 2007; Spetzler *et al.* 2008; Chadwick *et al.* 2010; Lumley, 2010; Pevzner *et al.* 2017; Yurikov *et al.* 2022), and geothermal reservoirs (Nibe & Matsushima 2021). Changes in fluid within the reservoir pores create significant reflection contrasts and cause delays in wave travel times through the reservoir. Additionally, seismic velocities are sensitive to stress, allowing time-lapse surveys to effectively image changes in seismic velocity caused by geomechanical deformation. Note that the approach to resource exploration and the assurance of integrity involve distinctly different considerations of risks and uncertainties. In recent years, the technology required for exploration has advanced considerably. For instance, the primary risks of a carbon dioxide Capture and Storage (CCS) project include $CO_2$ leakage and induced seismicity. Although there are numerous technology portfolios related to CCS monitoring, each varying in cost and benefit, a definitive methodology for CCS monitoring has yet to be established.

To ensure the long-term safe containment of injected $CO_2$ in reservoirs, and to acceptably manage the geomechanical deformation of the reservoir and caprock, several concerns must be addressed. First, with the current technology, our capabilities are limited to identifying $CO_2$ plumes; quantifying $CO_2$ saturation in the reservoir remains beyond the reach of existing geophysical monitoring techniques. Second, geomechanical deformation resulting from $CO_2$ injection can generate or reactivate fractures in the caprock, creating potential pathways for $CO_2$ leakage. Additionally, this deformation may induce seismicity, potentially compromising the integrity of both the reservoir and the caprock. However, most geomechanical parameters cannot be directly measured using existing geophysical exploration methods.

In evaluating gas saturation levels, it is well understood that seismic waves are highly sensitive to the presence of gas, yet they exhibit limited sensitivity to variations in the degree of gas saturation. Historically, the challenge has been the inability to differentiate between 'low gas' and 'high gas' saturations—a problem originating from the minimal responsiveness of *P*-wave and *S*-wave velocities to changes in gas saturation. Specifically, the presence of even a small amount of gas causes a sharp decrease in *P*-wave velocity, which then increases gradually with further increments in gas saturation. Conversely, *S*-wave velocity tends to increase as gas saturation rises. In the exploration of gas-bearing strata, the seismic method is predominantly employed. However, this method struggles to differentiate between low gas saturation (less than 5 per cent) and high gas saturation (approximately 30 per cent or more). This difficulty arises because the variation in seismic wave velocity between these two states is minimal. Seismic wave velocity is a parameter that integrates elastic constants and density, which is influenced by the interplay between these elastic constants and density.

In the assessment of geomechanical parameters using seismic methods, Poisson's ratio is evaluated based on the velocities of *P* waves and *S* waves, independent of density. However, the estimation of other parameters such as bulk modulus, shear modulus, and Young's modulus necessitates information about density. The development and application of these geomechanical parameters are subjects of ongoing research. As widely recognised in relevant literature, the production of conventional oil and natural gas alters the stress and strain fields both within and surrounding the reservoir due to a reduction in pore pressure, which consequently causes deformation (Herwanger & Horne 2009).

For instance, Guilbot & Smith (2002) observed a phenomenon in the North Sea where the reservoir composed of limestone contracts and the overlying strata expand during oil production, as detected by time-lapse seismic reflection surveys. From the perspective of reservoir management, recent research have significantly focused on reservoir geomechanical modelling that accounts for rock deformation. This approach complements traditional reservoir simulations, which primarily forecast the behaviour of fluids within the pore spaces (e.g. Herwanger & Koutsabeloulis 2011). Additionally, this modelling is used to quantitatively assess the geodynamic changes in the reservoir by correlating the temporal strain in the reservoir with changes in the position of the seismic reflection surface (Hatchell & Bourne 2005; Staples *et al.* 2007).

Furthermore, the orientation of fractures induced by hydraulic stimulation in a reservoir is contingent upon the direction of the principal stress. This directional stress varies with the production of oil and gas (Herwanger *et al.* 2013). Geomechanical approaches to reservoir management are critical for evaluating long-term well integrity and optimizing the trajectories of wells in oil and gas production (Herwanger *et al.* 2013). However, distinguishing between elastic constants and density information within seismic survey data presents significant challenges.

Muography is a non-destructive technique (Tanaka *et al.* 2023) that utilizes cosmic-ray muons to estimate the density distribution within an object. As these muons are generated by interactions between cosmic rays and atmospheric nuclei, they exhibit energy levels ranging from several GeV to over several TeV, enabling them to penetrate rock formations thicker than 1000 m. The transmission of muons through an object is inversely proportional to the integrated density along their path. Consequently, by counting the muons that pass through an object, its internal density distribution can be inferred.

The use of cosmic-ray muons to estimate underground density was first applied in civil engineering by George (1955), who measured the thickness of an overlying bedrock inside an underground tunnel, although the immense size of the measurement apparatus forced the experiment to be conducted at a laboratory scale within this tunnel. As the portability of the measurement apparatus improved, researchers such as Alvarez et al. (1970) started to explore hidden chambers within pyramids. Subsequent studies by Tarkova et al. (1973) and Bondarenko et al. (1974) aimed to estimate the density distribution of geological structures.Later, Malmqvist et al. (1979) explored the application of this technique for the physical prospecting of mineral resources.

Advancements in particle detector technology have not only led to more compact designs but also expanded their applications to include monitoring spatiotemporal variations within volcanoes (Nagamine 1995; Tanaka et al., 2005, 2007).In recent years, the feasibility of installing muon detectors in boreholes has been recognized, enhancing the potential for monitoring CCS through numerical simulations (Kudryavtsev et al., 2012; Klinger et al., 2015).Mellors *et al.* (2016) noted that integrating muon measurements with seismic data can improve density estimates and facilitate the resolution of elastic parameters through preliminary 1D modelling. Excluding Imano *et al.* (1999), there are limited examples of laboratory-scale experimental studies on cosmic-ray muon detection. Imano *et al.* (1999) successfully demonstrated the use of cosmic rays to measure soil moisture at the laboratory scale.

In this study, we extend the preliminary modelling and theoretical insights of Mellors *et al.* (2016) by proposing a coupled analysis of cosmic-ray muon detection and seismic exploration. This approach aims to differentiate the *P*- and *S*-wave velocities into two elastic constants—bulk and shear moduli—and densities, thus enabling a quantitative estimation of gas saturation and geomechanical parameters. Initially, we demonstrated the benefits of decomposing seismic wave velocity into density and two elastic constants through fluid substitution modelling. Subsequently, to assess the applicability and performance of the proposed method, we conducted laboratory-scale experiments using a combination of cosmic-ray muons and seismic waves.

## 2. METHOD

### 2.1 Separation of seismic velocity into elastic constants and density

When assuming a linear isotropic elastic material, the *P*-wave velocity ($V_P$) and *S*-wave velocity ($V_S$) are respectively expressed by eqs (1) and (2).

$$V_P = \sqrt{\frac{K + 4G/3}{\rho}} \tag{1}$$

$$V_S = \sqrt{\frac{G}{\rho}} \tag{2}$$

In eqs. (1) and (2), $K$, $G$ and $\rho$ denote the bulk modulus, shear modulus and density, respectively. These equations reveal that the velocities of $P$ waves and $S$ waves are functions of the elastic constants and density, illustrating a trade-off relationship among velocity data, elastic constants and density. Although the values of $P$-wave and $S$-wave velocities can be observed, it is impossible to uniquely determine $K$, $G$ and $\rho$ from eqs. (1) and (2) alone, as these equations present two equations with three unknowns. Although Poisson's ratio can be calculated from the $P$- and $S$-wave velocities independently of density, determining the density is essential for estimating other mechanical parameters.

Combining cosmic-ray muon detection with seismic exploration facilitates the differentiation of seismic wave velocity into two components: elastic constants and density. However, obtaining $S$-wave data proves more difficult than obtaining $P$-wave data. In these instances, the $P$-wave modulus ($M$) can be determined by integrating cosmic-ray muon data with $P$-wave data, as described in eq. (3). The $P$-wave modulus depends on factors such as lithology or rock type, effective stress, grain contact areas and occasionally pore fluid (Ikwuakor 2023).

$$V_P = \sqrt{\frac{K + 4G/3}{\rho}} = \sqrt{\frac{M}{\rho}} \tag{3}$$

In this study, numerical investigations were conducted on porous media using Gassmann's model (Gassmann 1951) to demonstrate the effectiveness of decomposing velocity information into density and elastic constants. Gassmann's equation, presented in eq. (4), relates the saturated bulk modulus of the rock to its porosity, as well as to the bulk moduli of the porous rock frame, mineral matrix and pore-filling fluids.

$$K_{\mathrm{sat}} = K^* + \frac{\left(1 - K^*/K_0\right)^2}{\phi/K_{\mathrm{f}} + (1-\phi)/K_{\mathrm{f}} + K^*/K_0^2}, \tag{4}$$

where $K_{\mathrm{sat}}$ denotes the saturated bulk modulus, $K_0$ indicates the bulk modulus of the mineral matrix, $K_{\mathrm{f}}$ represents the bulk modulus of the pore fluid, $K^*$ denotes the bulk modulus of the porous rock frame and $\phi$ indicates porosity. Furthermore, the relationship between the bulk modulus of formation water ($K_w$), gas saturation ($S_{\mathrm{g}}$), bulk modulus of gas ($K_{\mathrm{g}}$) and bulk modulus of the pore fluid ($\rho_{\mathrm{f}}$) is described in eq. (5).

$$K_{\mathrm{f}} = \frac{1}{(1 - S_{\mathrm{g}})/K_w + S_{\mathrm{g}}/K_{\mathrm{g}}} \tag{5}$$

However, as the shear modulus of the dry rock is unaffected by the fluid within the pores, eq. (6) holds true.

$$G_{\mathrm{sat}} = G_{\mathrm{dry}}, \tag{6}$$

where $G_{\mathrm{sat}}$ and $G_{\mathrm{dry}}$ denote the saturated and non-saturated shear modulus, respectively. The relationship between the fluid density ($\rho_{\mathrm{f}}$), porosity ($\phi$), the density of the rock matrix ($\rho_{\mathrm{m}}$) and the bulk density ($\rho_{\mathrm{b}}$) is described in eq. (7).

$$\rho_{\mathrm{b}} = \rho_{\mathrm{m}}(1 - \phi) + \rho_{\mathrm{f}}\phi \tag{7}$$

Furthermore, the relationship between the formation water density ($\rho_w$), gas saturation ($S_{\mathrm{g}}$), gas density ($\rho_{\mathrm{g}}$), porosity ($\phi$) and fluid density ($\rho_{\mathrm{f}}$) is described in eq. (8).

$$\rho_{\mathrm{f}} = \rho_{\mathrm{w}}(1 - S_{\mathrm{g}}) + \rho_{\mathrm{g}} S_{\mathrm{g}} \tag{8}$$

Accordingly, if the porosity ($\phi$) is maintained constant and the gas saturation ($S_{\mathrm{g}}$) is varied, the changes in the $P$-wave velocity ($V_P$), $S$-wave velocity ($V_S$), saturated bulk modulus ($K_{\mathrm{sat}}$), saturated shear modulus ($G_{\mathrm{sat}}$), $P$-wave modulus ($M$) and bulk density ($\rho_{\mathrm{b}}$) can be determined. The parameters used in these calculations are listed in Table 1. The physical properties shown in Table 1 are not based on a specific region but rather vary with two types of porosity (0.2 and 0.3) and gas saturation (from 0 to 0.9) in consolidated sandstone.

### 2.2 Muon observation at a laboratory scale

We developed a prototype muon measurement system at the laboratory scale, as depicted in Fig. 1. The system comprises three plastic scintillators. The vertical separation between the top and the second scintillator is 10 cm, whereas the separation between the second and the bottom scintillator is 140 cm. Each scintillator measures 20 cm × 20 cm and exhibits a thickness of 2 cm. Detection of a muon occurs when it triggers a reaction in the scintillators, and the system counts the number of muons passing through all three scintillators simultaneously. To attenuate cosmic-ray electrons, we placed a 3-cm-thick lead plate between the top and second scintillators. The energy absorption of muon particles is a function of the product of the penetration length and the density of the material, referred to as the density length.

By counting the number of muons that penetrate a material and assuming they travel in a straight path, we can estimate the density of the material, provided its length is known. Typically, the observed muon count is converted into a density measurement using empirical formulas, such as Miyake's equation (Miyake 1979). However, this standard method is not applicable when measuring muons within a building due to the influence of the building's structure and the presence of heavy objects on the floors above the observation site. Therefore, the first step should determine the relationship between the density length and the muon count rate in such environments.

Using the prototype measurement system, we assessed the variations in water depth and muon flux within a water tank. By plotting the muon flux at various depths, we established a linear approximation of the relationship between density length and muon flux. Considering the density of water to be 1 g cm$^{-3}$, the water depth directly corresponded to the density length. The dimensions of the tank were 40 cm deep, 60 cm wide, and 50 cm long. The water depth in the tank was adjusted from 0 to 40 cm in 10 cm increments, with three 24-hr measurement sessions conducted at each depth increment.

We computed the hourly muon counts and determined the average value and standard deviation on three separate occasions. We should note that the observed average counts across the three measurements can be fluctuated due to various meteorological factors, primarily barometric pressure and temperature changes (Maghrabi *et al.* 2015). We then graphed the correlation between muon counts per hour over a 24-hr period and water depth, employing a linear approximation to formulate an equation. This equation accounts for the specific characteristics of the building housing the laboratory and the direction from which muons arrive at a particular location within the building. Utilizing this equation, the density length can be estimated based on the observed number of muons per hour.

Muon particles can penetrate not only vertically (zenith angle = 0°) but also at oblique angles, leading to varying penetration lengths. In this study, we simulated the geometric configuration of muon detectors as illustrated in Fig. 1. We generated 100 000 muon particles and tracked their detection by the muon detectors to calculate the average zenith angle. The calculated average zenith angle was approximately 6.1°. This value enabled us to adjust the density length by dividing it with the cosine of the zenith angle relative to the vertical length of the target.

**Table 1.** Parameters used in the fluid substitution approach based on Gassmann's model.

| Parameter and unit | Value |
|---|---|
| Bulk modulus of the mineral matrix ($K_0$) (GPa) | 38.0 |
| Bulk modulus of the porous rock frame ($K^*$) (GPa) | 15.0 |
| Bulk modulus of formation water ($K_w$) (GPa) | 2.38 |
| Bulk modulus of gas ($K_g$) (GPa) | 0.0021 |
| Shear modulus of dry rock ($G_{dry}$) (GPa) | 15.0 |
| Density of the rock matrix ($\rho_m$) (g cm$^{-3}$) | 2.62 |
| Density of formation water ($\rho_w$) (g cm$^{-3}$) | 1.0 |
| Density of gas ($\rho_g$) (g cm$^{-3}$) | 0.0101 |
| Porosity ($\phi$) | 0.2, 0.3 |
| Gas saturation ($S_g$) | Changing in increments of 0.01 from 0.0 to 0.9 |

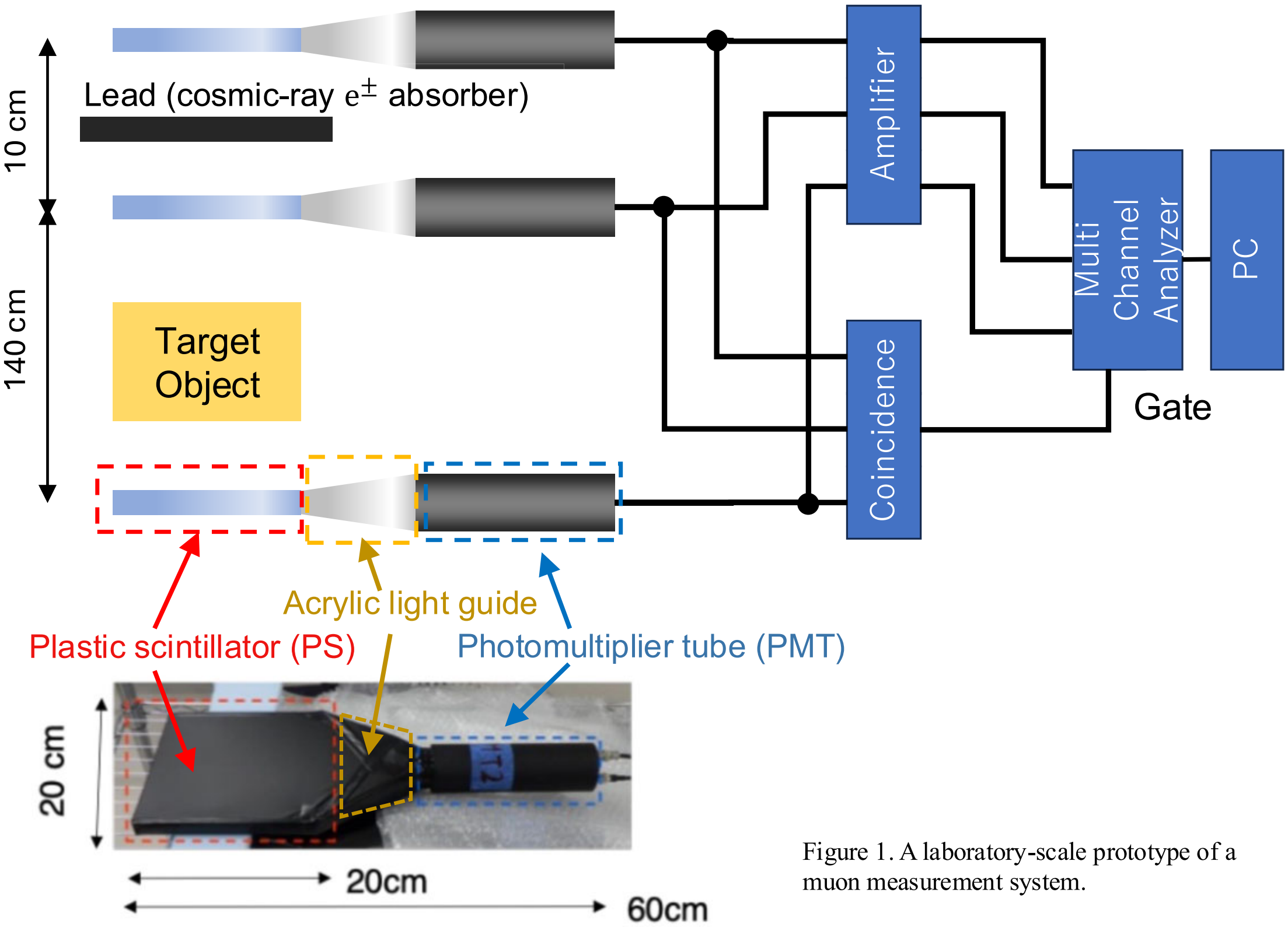

**Figure 1.** A laboratory-scale prototype of a muon measurement system.

A laboratory experiment was conducted on the fourth floor of a seven-story building to assess the feasibility of using laboratory experiments to measure changes in water depth and variations in muon flux within a water tank. To this end, muon particles were numerically simulated using GEANT4 (Agostinelli *et al.* 2003), a software designed for simulating the interactions between particles and matter. We assumed that the ceiling of each floor in the building was composed of 30 cm of concrete with a density of 2.4 g cm$^{-3}$. The energy of the muons in the simulation ranged from 100 MeV to 1000 GeV.

Fig. 2 illustrates the penetration of 100 000 muons, each with energies ranging from 600 to 1200 MeV, through a seven-story building when incident vertically. In this figure, yellow lines represent coordinates, green lines depict gamma rays produced in the concrete, and red lines represent muons or electrons. The red and yellow lines are nearly indistinguishable due to their similarity. The selected energy range of 600–1200 MeV corresponds to the minimum energy required for muons to penetrate standard rock, as determined by experimental results reported by the Particle Data Group.

Within this energy range, the muon flux intensity (cm$^2$ s$^{-1}$) measured on the second floor was 4.18e−06. The muon count was recorded hourly over a 24-hr period on three separate occasions

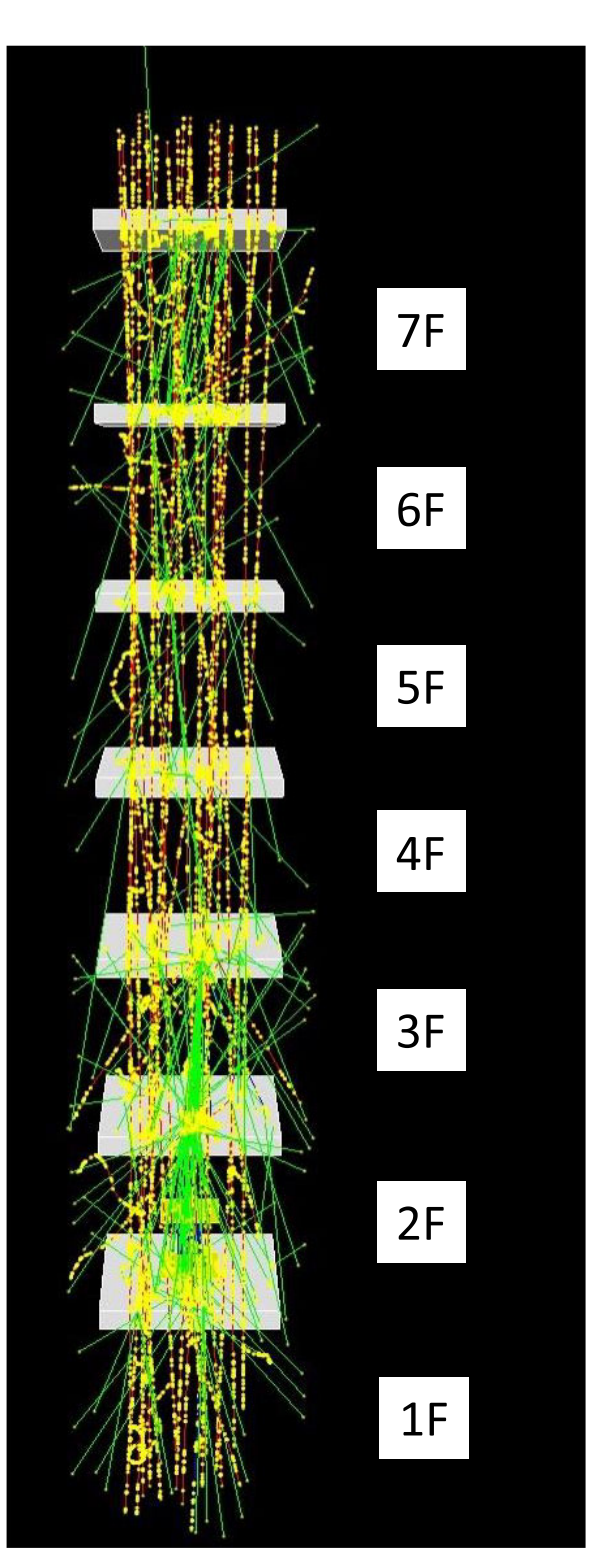

**Figure 2.** Simulation of 100 000 vertically incident muons with energies ranging from 600 to 1200 MeV penetrating a seven-story building.

using acrylic and aluminium blocks, each measuring 20 cm × 20 cm × 20 cm, as depicted in Fig. 3.

### 2.3 Ultrasonic measurement

An ultrasonic laboratory experiment was conducted on acrylic and aluminium blocks using piezoelectric transducers manufactured by Panametrics (model number: V103-RM), as outlined with reference to Matsushima *et al.* (2008). A high-voltage electric pulse activated the piezoelectric crystal, generating an ultrasound pulse through the piezoelectric effect. The transducer, specifically designed for *P*-wave generation, produced a one-cycle sinusoidal wave at 1 MHz. This wave was transmitted through the sample and received by a similar transducer with a central frequency of 1 MHz. The received waveform signals were amplified and recorded on a digital oscilloscope, which converted the signals from analogue to digital.

Synchronization between the trigger signal of the pulse generator and the oscilloscope enabled precise determination of traveltime information. To improve the signal-to-noise ratio, each signal was averaged 512 times. Matsushima *et al.* (2008) calculated the coefficient of variation (CV) for the reproducibility of velocities, average frequencies and peak-to-peak amplitudes of the observed waveforms as 0.36 per cent, 0.30 per cent and 1.94 per cent, respectively. These CV values indicate that the ultrasonic waveform measurement system demonstrated high stability and repeatability with considerable accuracy.

Ultrasonic measurements were performed using two piezoelectric elements configured for *P*-wave generation, as illustrated in Fig. 4. Both elements were set with oscillation and incident angles at 45°. This configuration facilitated the simultaneous recording of *P*- and *S*-wave waveforms due to the radiation pattern of the piezoelectric elements. The source and receiver were placed at 15 sequential locations, spaced 1 cm apart, beginning 2 cm from the sample edge. By identifying the first arrivals in the recorded waveforms of both *P* and *S* waves, a plot of traveltime against propagation distance was created. Based on this plot, the velocities of the *P* and *S* waves were calculated based on the slope.

## 3. RESULTS

Figs 5(a)–(f) illustrate the variations in *P*-wave velocity ($V_P$), *S*-wave velocity ($V_S$), saturated bulk modulus ($K_{\mathrm{sat}}$), saturated shear modulus ($G_{\mathrm{sat}}$), *P*-wave modulus ($M$) and bulk density ($\rho_{\mathrm{b}}$) as functions of gas saturation ($S_{\mathrm{g}}$) for two different porosities (the solid line represents porosity = 0.3, and the dashed line represents porosity = 0.2). Fig. 5(a) demonstrates that with a minimal gas presence, $V_P$ decreases sharply, followed by a gradual increase as gas saturation continues to rise. Conversely, Fig. 5(b) shows that $V_S$ increases with increasing gas saturation. The behaviours of $V_P$ and $V_S$ are influenced by the interplay between density and elastic constants.

In Fig. 5(c), the saturated bulk modulus decreases sharply with a small amount of gas and then stabilizes with further increases in gas saturation. Conversely, the saturated shear modulus does not depend on gas saturation, as illustrated in Fig. 5(d). This independence arises because the shear modulus of dry rock is unaffected by the presence of fluid within the pores. Fig. 5(e) demonstrates that the *P*-wave modulus sharply decreases before stabilizing with increased gas saturation. Additionally, Fig. 5(f) displays that the bulk density decreases linearly as gas saturation increases.

Fig. 6 presents the average muon counts per hour from three repeated measurements at five different water depths, with standard deviations shown as error bars. To minimize visual clutter from overlapping error bars, single-sided error bars are used. The figure indicates that up to approximately 12 h, fluctuations were observed in the average muon counts; however, these values stabilize after this period. As described above, the observed fluctuations in the average counts across the three measurements can be attributed to various meteorological factors, primarily barometric pressure and temperature changes.

Our three repeated laboratory measurements indicated that the average muon counts correlated with water depth: higher counts were observed at shallower depths and lower counts at greater depths. However, this relationship was not linear, as the differences in muon counts did not consistently correspond to uniform 10 cm intervals of water depth. We plotted the muon counts for five different averaged water depths, including their standard deviations and conducted a linear regression analysis to approximate these relationships with a straight line (Fig. 7). In Fig. 7, the horizontal axis denotes the water depth (equivalent to the density length), and the vertical axis represents the muon counts. In this linear fitting, we calculated the reduced chi-squared value and obtained 1.688. A reduced chi-squared value of 1.688 indicates that indicates that the model is not perfectly fitting the data but is not dramatically off either. We need to refine the model or check the data and error estimates to improve the fit.

The derived approximation line is an expression that accounts for the characteristics of the building housing the laboratory and the direction of muon arrival at a specific location within the building. Our numerical muon simulation in Fig. 2 indicates that low-energy muons can be used to differentiate between various water depths in the tank. We conducted three 24-hr measurements of the muon count in both acrylic and aluminium blocks, as illustrated in Fig. 3. Subsequently, we calculated the average and standard deviation of

Acrylic

Aluminum

**Figure 3.** Muon count measurement for two distinct targets: (a) acrylic block and (b) aluminium block (20 cm × 20 cm × 20 cm).

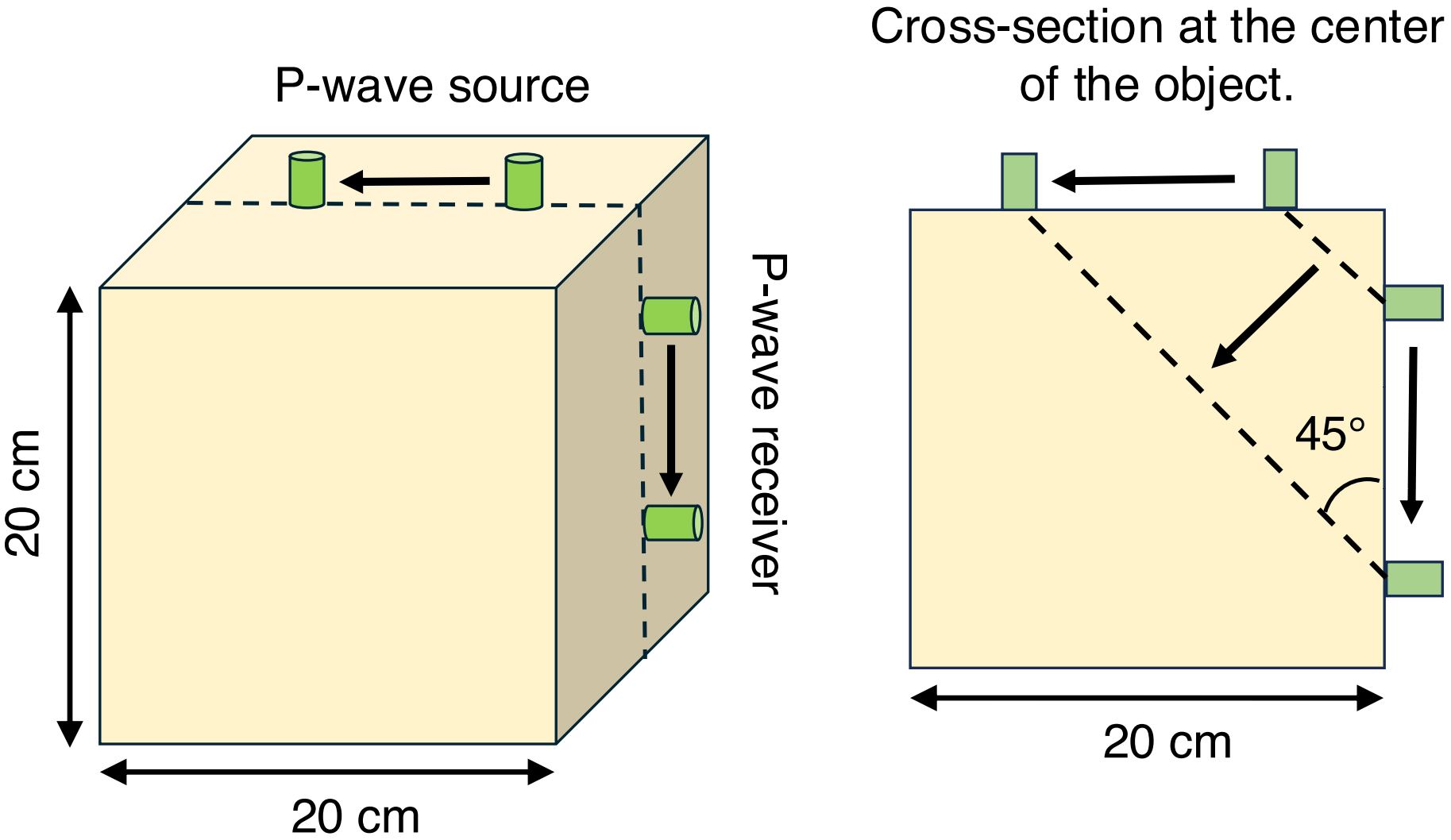

**Figure 4.** Ultrasonic measurements were conducted by positioning two piezoelectric elements specifically designed for *P*-wave generation, with both the oscillation and incident angles set to 45°.

the parameters derived from these muon counts per hour, which are presented in Table 2. Similarly, we calculated the average and standard deviation of the parameters for four water depths (10, 20, 30 and 40 cm) in Table 3.

Fig. 8 presents the ultrasonic data for *P* waves and *S* waves observed in two different targets: (a) an acrylic block and (b) an aluminium block. The relative amplitudes are maintained in each panel. In this figure, the locations of the *P*-wave and *S*-wave events are prominently marked by blue and red dashed lines, respectively. The identification of *S*-wave events was verified through additional experiments that analysed the radiation pattern of the piezoelectric elements, confirming that these were not surface waves. We identified the first arrivals in the recorded waveforms for both *P* waves and *S* waves, and subsequently plotted these on a graph showing traveltime versus propagation distance.

We calculated the *P*- and *S*-wave velocities using the slopes derived from the acrylic and aluminium blocks, as outlined in Table 4. The experimental velocities of the *P* waves and *S* waves are compared with those from catalogued values, demonstrating a general concordance. Utilizing the average density values from muon detection (Table 2) and the velocity values for *P* waves and *S* waves from ultrasonic experiments (Table 4), we subsequently determined

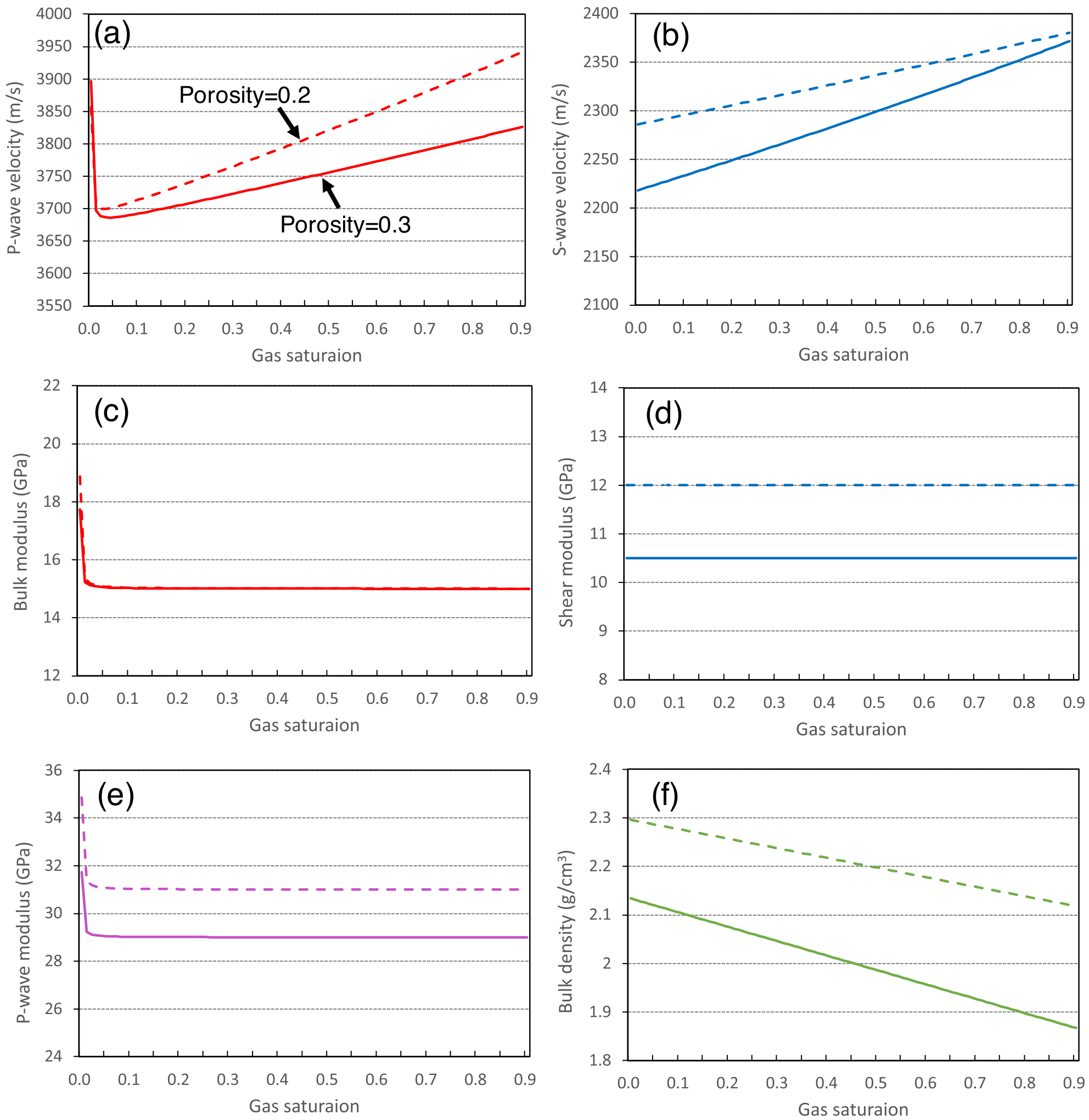

**Figure 5.** Variations in parameters with respect to gas saturation: (a) *P*-wave velocity, (b) *S*-wave velocity, (c) saturated bulk modulus, (d) saturated shear modulus, (e) *P*-wave modulus and (f) bulk density. The solid line represents porosity = 0.3, and the dashed line represents porosity = 0.2.

the bulk and shear moduli for the acrylic and aluminium blocks, presented in Table 5.

## 4. DISCUSSION

Fig. 5 illustrates the benefits of decomposing seismic wave velocity into density and two elastic constants through fluid substitution modelling. The behaviour of *P*- and *S*-wave velocities is influenced by the interplay between density and these elastic constants. Upon this decomposition, our analysis revealed a linear decrease in bulk density with increasing gas saturation, as depicted in Fig. 5(f). This finding suggests that density measurements can effectively predict pore fillings. However, it is important to note that when calculating gas saturation, it is necessary to determine porosity. Density is influenced by both gas saturation and porosity (Fig. 5f), while the bulk modulus is influenced by gas saturation but is not sensitive to porosity (Fig. 5c). Conversely, the shear modulus is affected solely by porosity (Fig. 5d). Therefore, it is conceivable to estimate porosity information from the shear modulus. Predicting the types and characteristics of pore fillings using seismic waves is challenging due to complex interactions during wave propagation, such as pore geometry effects, wave-induced fluid flow, gas bubble oscillation and the impact of gas bubble size. However, these complex interactions do not compromise the reliability of density information, thereby facilitating accurate predictions of the types and characteristics of pore fillings.

Van Koughnet *et al.* (2003) determined from well logging data that density measurements are more effective than Amplitude Versus Offset (AVO) characteristics derived from seismic data for identifying pore fillings. Additionally, for reservoir simulations that integrate fluid flow and geomechanics, two elastic constants and density—geomechanical parameters—are essential. These parameters enhance the precision of monitoring and prediction activities. However, note that the geomechanical parameters acquired from seismic wave data are dynamic and must be converted to static parameters before they can be used in reservoir geomechanics simulations.

To verify the applicability and accuracy of the proposed method for separating seismic wave velocity into two types of elastic constants and density, muon and ultrasonic data were collected from the same target in laboratory experiments. The average muon counts per hour, derived from three repeated measurements at each of five water depths (refer to Fig. 6), exhibited fluctuations throughout the measurement period. These fluctuations are primarily attributable to meteorological influences, particularly barometric pressure and temperature variations. To mitigate these meteorological effects, Imano *et al.* (1999) employed a reference measurement instrument alongside the primary instrument to correct the observed muon flux. A future objective will be to implement this correction method to improve the accuracy of muon measurements.

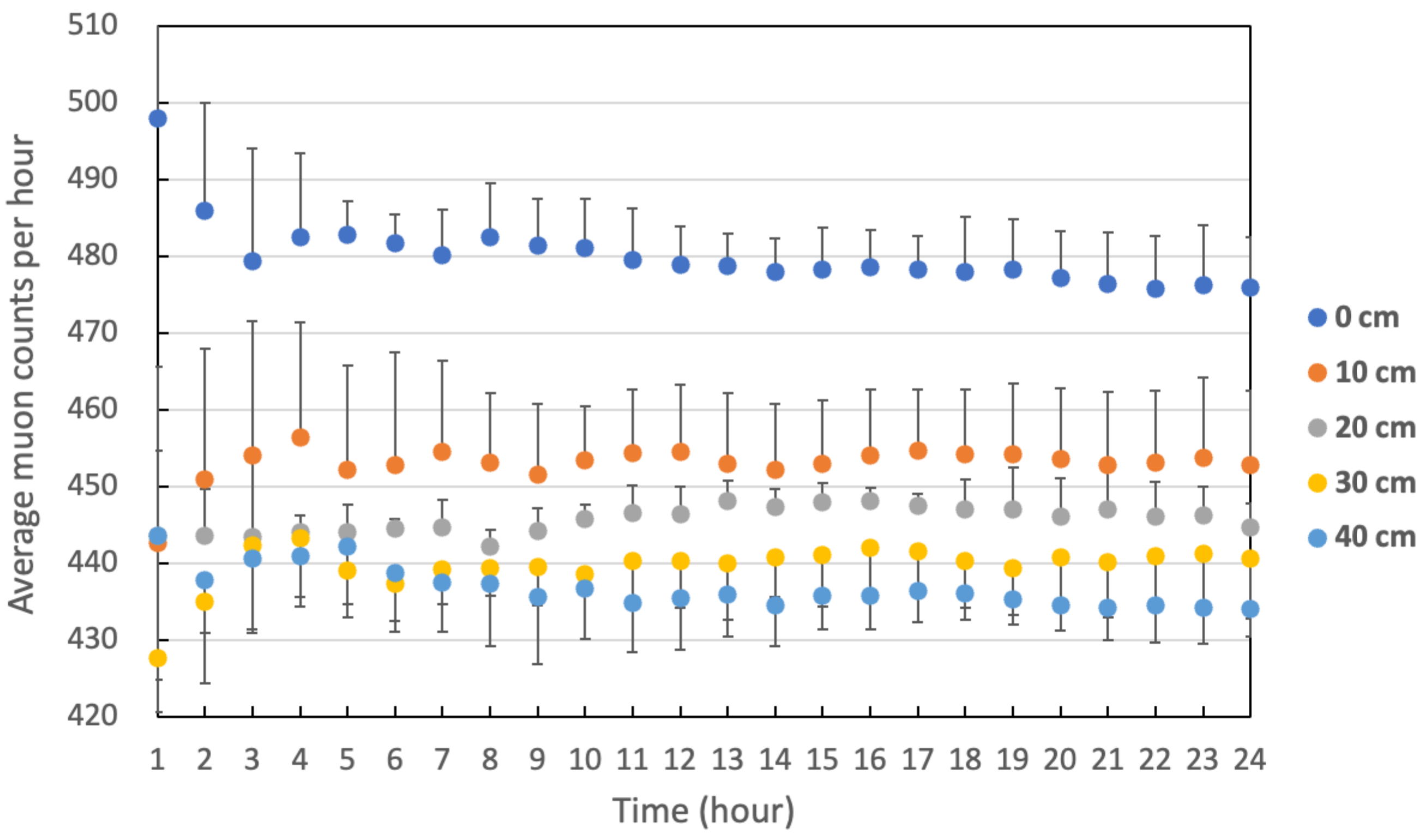

**Figure 6.** Hourly averages of muon counts from three repeated measurements for each of the five different water depths are shown, with their standard deviations indicated by single-sided error bars.

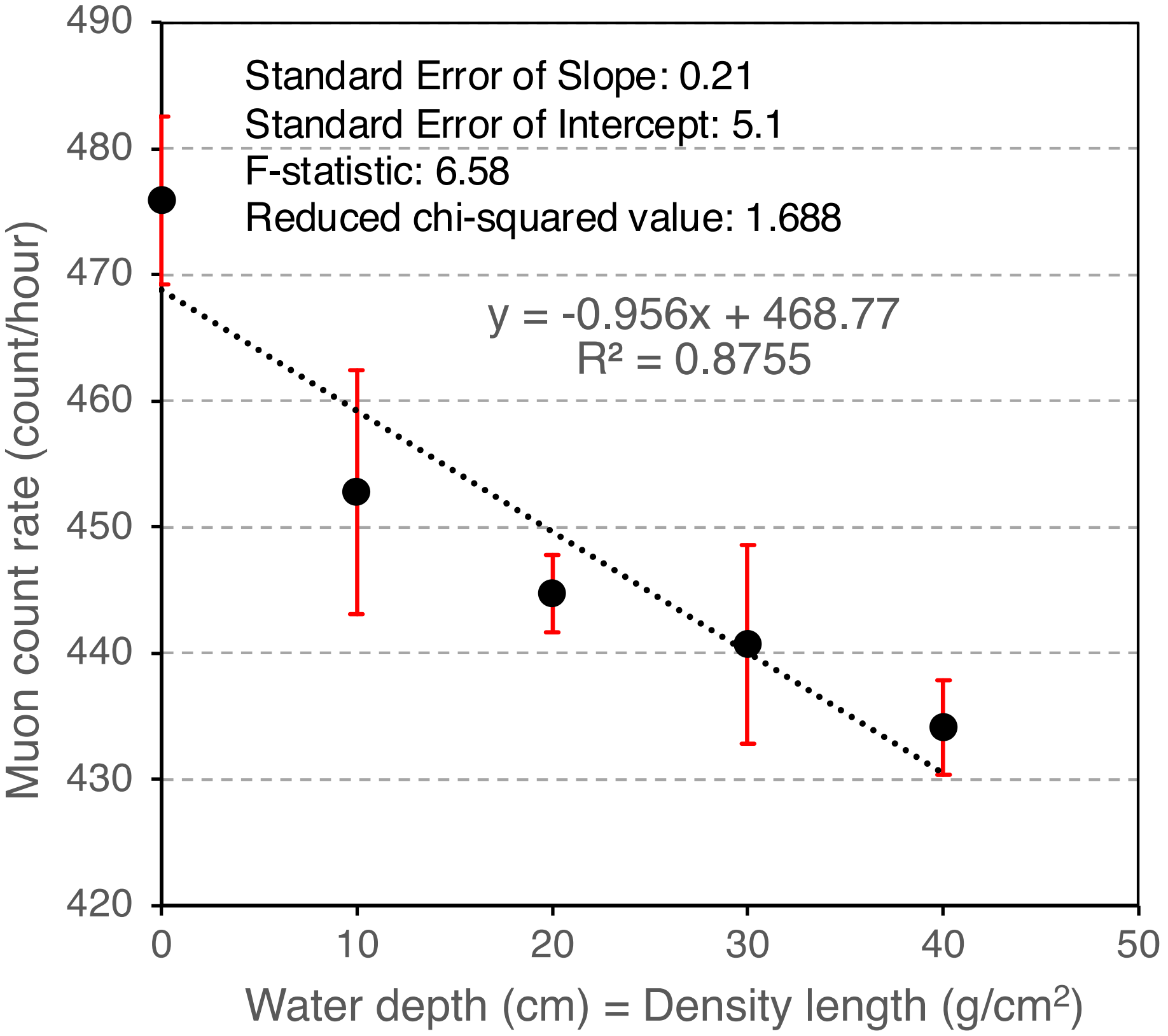

**Figure 7.** Muon counts for five different averaged water depths, including their standard deviations indiacted by error bars, with a linear regression analysis applied for a straight-line approximation. The errors on the fit parameters are also shown in this graph.

**Table 2.** Average value and standard deviation of parameters obtained from measuring muon counts per hour over 24 hr, repeated three times in acrylic and aluminium blocks. Catalogue values of densities are provided as well.

| Parameter | Acrylic | Aluminium |
|---|---|---|
| Muon counts per hour | 444.6 ± 5.6 | 419.4 ± 2.3 |
| Density length (g cm$^{-2}$) | 25.3 ± 5.9 | 51.6 ± 2.4 |
| Estimated Density (g cm$^{-3}$) | 1.27 ± 0.29 | 2.58 ± 0.12 |
| True Density (g cm$^{-3}$) | 1.2 | 2.7 |
| Error (per cent) | −18.3 ∼ 30.0 per cent | 0 ∼ 8.8 per cent |

In Fig. 7, we derived the relationship between muon flux and density length, enabling the conversion of muon flux measurements into density length. This relationship accounts for the specific characteristics of the building housing the laboratory and the direction from which muons arrive at a particular location within the building. However, if there is a change in the position of the measuring instrument or an alteration in the size of the muon detector, it will be necessary to rederive this relationship. Furthermore, owing to potential repositioning of instruments or furniture on upper floors, this relationship should be periodically recalculated.

We determined the density of two different materials, an acrylic block and an aluminium block, using muon transmission. The average density and standard deviation were calculated from three measurements for each material (refer to Table 2). The results indicated that the error associated with the density measurement of the aluminium block was smaller than that of the acrylic block. This finding suggests that denser materials (larger density lengths) exhibit smaller errors in density measurements. The trend is also evident in the measurement results for different water depths shown in Table 3. The underlying reason is that as the density of the material increases, fewer muons pass through, leading to reduced fluctuations in the measurement data.

In this study, we highlight the significance of decomposing seismic velocity data into density and two elastic constants. The challenges associated with accurately estimating density and elastic constants are well acknowledged. Efforts to address these challenges through advancements in muography and analytical technologies are underway. Additionally, the estimation of density and elastic constants at the reservoir scale is gaining attention. Currently, density information can be obtained through two primary methods: gravity and seismic exploration. We will discuss these methods, their challenges and other relevant aspects below.

The gravity exploration method is utilized to determine the density of subsurface rocks and is recognized as an effective reconnaissance tool for identifying large-scale geological structures. Its primary function is to describe sedimentary basins by detecting features such as faults, intrusive bodies and anticlinal structures. With advancements in measurement technology, this method has increasingly been applied to assess reservoir scales in recent years (Elliott & Braun 2017). Since the 2000s, gravity monitoring has been employed in the surveillance of oil and gas reservoirs. Challenges such as achieving high precision (1 $\mu$gal or less) and enhancing portability have been addressed by the commercialization of superconducting gravimeters, notably the iGrav from GWR Instruments, which have significantly contributed to the field (Hinderer *et al.* 2015).

Specifically, researchers have emphasized on monitoring the application of the Steam-Assisted Gravity Drainage (SAGD) method in unconventional oil resources. Oil sands contain a type of petroleum known as bitumen, which has become denser due to the movement of oil reservoirs near the surface, influenced by erosion, crustal movements, and the loss of lighter components through volatilization. The SAGD method involves injecting steam underground to decrease the viscosity of heavy oil, thereby facilitating its recovery. The area occupied by the injected steam is referred to as a steam chamber. Monitoring the spatial and temporal expansion of the steam chamber is crucial for optimizing economic benefits, such as identifying areas for bitumen recovery, and for mitigating environmental risks, such as detecting bitumen leaks beyond the designated area.

Reitz *et al.* (2015) demonstrated the capability of detecting water vapour circulation regions using time-varying gravity gradiometry exploration and the gravity anomaly method. Additionally, they described the feasibility of monitoring bitumen recovery regions with the gravity anomaly method. Elliott & Braun (2017) further substantiated these findings using a numerical model. However, the gravity method, which utilizes a potential field, exhibits a trade-off between exploration depth and spatial resolution. Consequently, as the depth of the exploration target increases, the resolution diminishes.

There are three primary methods for deriving density information from seismic data: employing an empirical density–velocity formula utilizing AVO analysis and implementing full-waveform inversion. One prominent approach involves Gardner's formula, which applies an empirical density–velocity relationship. Gardner *et al.* (1974) formulated empirical equations based on density and sonic logging data collected from sedimentary basins of various ages and depths. This study established linear relationships between several rock types.

Gardner's equation generally provides accurate approximations for sandstone, shale and carbonate rocks; however, its applicability to coal formations and evaporites is limited (Quijada & Stewart 2007). Furthermore, the equation often yields significant errors, even when the approximation appears relatively accurate, rendering it unsuitable for quantitative analyses that require detailed consideration of specific rock characteristics, such as reservoir analysis, geomechanical analysis and porosity pressure analysis (Nwozor *et al.* 2017). Although Lindseth (1979) introduced an alternative method using an empirical formula to describe the density–velocity relationship, it did not represent a substantial improvement.

Numerous studies have indicated that the results from AVO analysis are highly unstable and generally ineffective for estimating density information (e.g. Behura *et al.* 2010; Schulte & Manthei 2014). To address this issue, seismic traces with the broadest possible range of offset distances can be employed, or alternatively, to use traces with relatively short offset distances when density contrast is the primary factor influencing amplitude changes (Li 2005). Nonetheless, obtaining reliable density information from actual field data remains unfeasible. Lines (1999) conducted a sensitivity analysis and quantitatively demonstrated the challenges associated with deriving density information from AVO analysis.

Numerous studies have highlighted the challenges associated with accurately estimating density information in methods that employ full-waveform inversion (Köhn et al., 2012; Alkhalifah & Plessix 2014). Owing to these challenges, various approaches are adopted in the application of full-waveform inversion, including assuming a homogeneous density, employing a velocity–density empirical relationship and determining the acoustic impedance. The difficulty in deriving density information using this method stems from two primary factors. First, there is an interdependent trade-off relationship between velocity and density. Second, the scattering patterns of radiation vary with changes in velocity and density.

**Table 3.** Average value and standard deviation of parameters obtained from measuring muon counts per hour over 24 hr, repeated three times in four water depths (10, 20, 30 and 40 cm).

| Parameter | Water depth (10 cm) | Water depth (20 cm) | Water depth (30 cm) | Water depth (40 cm) |
|---|---|---|---|---|
| Muon counts per hour | 452.8 ± 9.7 | 444.7 ± 3.1 | 440.7 ± 7.9 | 434.1 ± 3.7 |
| Density length (g cm$^{-2}$) | 16.7 ± 10.1 | 24.9 ± 3.2 | 29.0 ± 8.2 | 36.4 ± 3.9 |
| Estimated density (g cm$^{-3}$) | 1.67± 1.01 | 0.62 ± 0.08 | 0.73 ± 0.21 | 0.91 ± 0.09 |
| True density (g cm$^{-3}$) | 1.0 | 1.0 | 1.0 | 1.0 |
| Error (per cent) | –33.9 ∼ 168.8 per cent | –45.9 ∼ –29.7 per cent | –47.9 ∼ –6.8 per cent | –18.9 ∼ 0.7 per cent |

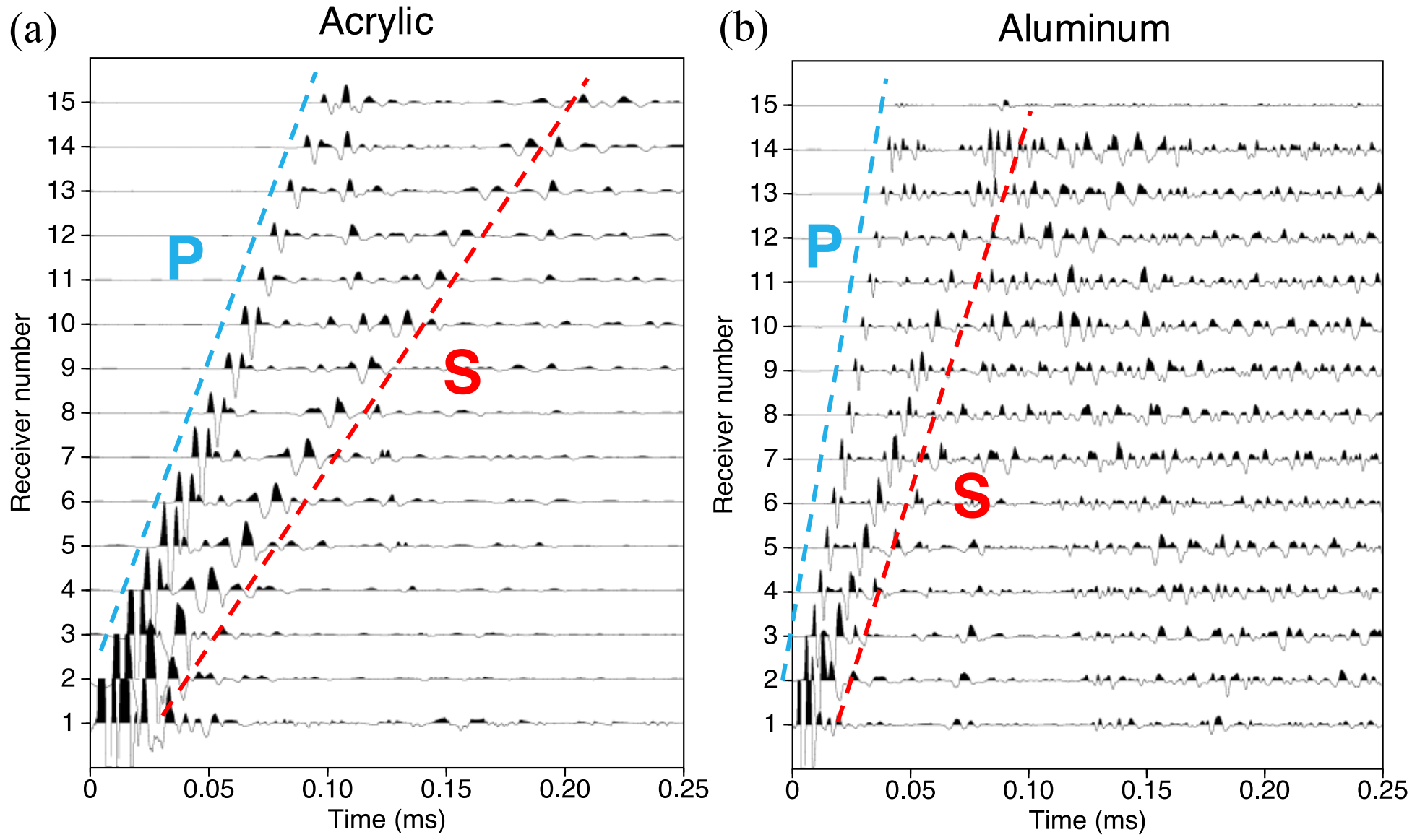

**Figure 8.** Observed ultrasonic data for *P* and *S* waves on two targets: (a) acrylic block and (b) aluminium block. Relative amplitude is maintained in each panel. Two dashed lines in each panel mark the positions of *P*-wave and *S*-wave events, respectively.

**Table 4.** Velocities of *P* waves and *S* waves determined from slopes for acrylic and aluminium blocks. Catalogue values of *P* waves and *S* waves are provided as well.

| Parameter | Acrylic | Aluminium |
|---|---|---|
| Vp (m s$^{-1}$) | 2721.5 | 6175.5 |
| Vs (m s$^{-1}$) | 1376.1 | 3068.8 |
| True Vp (m s$^{-1}$) | 2730 | 6320 |
| True Vs (m s$^{-1}$) | 1360–1400 | 3240 |

**Table 5.** Bulk modulus and shear modulus determined from the average density values obtained from muon detection and the velocity values of *P* waves and *S* waves obtained from ultrasonic experiments for acrylic and aluminum blocks. True values of bulk and shear modulus calculated from catalogue values of densities, *P* waves and *S* waves are provided as well.

| Parameter | Acrylic | Aluminum |
|---|---|---|
| Estimated bulk modulus (GPa) | 6.18 | 65.99 |
| True bulk modulus (GPa) | 5.81–5.98 | 69.07 |
| Estimated shear modulus (GPa) | 2.39 | 24.29 |
| True shear modulus (GPa) | 2.22–2.35 | 25.43 |

Luo & Wu (2018) demonstrated that the radiation pattern of scattered waves varies between velocity and density fields when Rayleigh scattering is assumed. Specifically, the radiation pattern in the velocity field is isotropic, allowing it to be observed at any offset distance. In contrast, the radiation pattern in the density field is confined to regions with large scattering angle, rendering it observable mainly near the offset. Luo & Wu (2018) leveraged these differences in scattered radiation characteristics to successfully derive a density distribution by applying waveform inversions to numerical simulation data.

This method has not yet been applied to real field data. Additionally, the quantitative assessment of the accuracy and resolution of density estimation methods using full-waveform inversions of seismic data remains underdeveloped. The commonly employed chequerboard method does not provide a quantitative measure of the reliability of analysis results; instead, it is used to qualitatively and visually assess reliability. However, it has been criticized for potentially misleadingly indicating high resolution in the analysis results when small-scale inhomogeneities are apparent (Fichtner & Trampert 2011). To address this issue, Fichtner & Trampert (2011) introduced a quantitative resolution evaluation method as an alternative to the chequerboard method and applied this new method to evaluate resolution at a crustal scale. Nonetheless, no such performance evaluations have been conducted at the scale of oil and natural gas reservoirs.

As discussed earlier, current geophysical exploration techniques for obtaining density information face challenges regarding resolution and accuracy. The resolution and accuracy of density measurements obtained through the proposed method should be compared with those derived from existing geophysical exploration techniques. Recent research has initiated numerical simulations of muon exploration at the reservoir scale under actual field conditions. Pieczonka *et al.* (2020) demonstrated the capability to detect density anomalies in an SAGD reservoir by modelling horizontal muon detector arrays within horizontal boreholes and analysing their responses through both forward and inverse modelling. They also suggested the potential to enhance vertical resolution by optimizing detector placement, such as employing sensors in vertical boreholes instead of solely at a single depth in a horizontal array. Consequently, future studies should extend this research to include a broader range of targets.

## 5. CONCLUSIONS

In this study, we highlight the utility of decomposing seismic wave velocities into two elastic constants and density for assessing pore inclusions and the geomechanical properties of the subsurface. Initially, the proposed Gassmann's model-based fluid substitution approach demonstrated that bulk density decreases linearly with increasing gas saturation. This finding highlights the benefit of distinguishing seismic wave velocity into density and two elastic constants. Furthermore, our modelling suggests that density data can effectively predict the types of pore fillings. Although the prediction of pore filling types and characteristics through seismic waves is complicated due to multiple interactions, the inclusion of density data simplifies these predictions. Second, we demonstrate the feasibility of measuring two elastic constants and density by employing a combination of cosmic-ray muons and seismic exploration at the laboratory scale. We highlighted the need to establish a relationship that converts muon flux into density length. This relationship considers the characteristics of the building housing the laboratory and the direction from which the muons arrive at a specific location within the building. To deploy this method at the reservoir scale under real field conditions in the future, the resolution and accuracy of this method need to be comparatively evaluated with those of existing geophysical exploration techniques such as gravity and seismic exploration through numerical simulations.

## DATA AVAILABILITY

Data associated with this research are available and can be obtained by contacting the corresponding authors.

## ACKNOWLEDGMENTS

We greatly acknowledge the thorough reviews and constructive comments of one anonymous reviewer, which helped increase the quality of the manuscript. This work was supported by the Mohammed bin Salman Center for Future Science and Technology for Saudi Arabia-Japan Vision 2030 at the University of Tokyo (MbSC2030). This study was supported by ERI JURP 2022-H-01 (Earthquake Research Institute, University of Tokyo).